\documentclass[Proceedings]{SciPost}
\usepackage{graphicx}
\usepackage{subcaption}

\usepackage{url}
\usepackage{hyperref}
\definecolor{nicered}{rgb}{0.5,0.,0.}
\definecolor{nicegreen}{rgb}{0.,0.5,0.}
\definecolor{niceblue}{rgb}{0.,0.,0.5}
\hypersetup{colorlinks,citecolor=nicegreen,linkcolor=nicered,urlcolor=niceblue}

\usepackage{amsmath,amssymb}
\usepackage{physics}
\DeclareSymbolFont{usualmathcal}{OMS}{cmsy}{m}{n}
\DeclareSymbolFontAlphabet{\mathcal}{usualmathcal}

\newcommand{\GeV}{\textrm{GeV}}
\newcommand{\TeV}{\textrm{TeV}}

\begin{document}

{\hfill MSUHEP-21-015, PITT-PACC-2116, SMU-HEP-21-11}

\begin{center}{\Large \textbf{
The photon content of the proton in the CT18 global analysis}
}\end{center}

\begin{center}
	Keping Xie\textsuperscript{1$\star$},
	T.~J.~Hobbs\textsuperscript{2,3,4,5},
	Tie-Jiun Hou\textsuperscript{6},
	Carl Schmidt\textsuperscript{7},
	Mengshi Yan\textsuperscript{8},
	and C.-P. Yuan\textsuperscript{7}
\end{center}

\begin{center}\small
	{\bf 1} Department of Physics and Astronomy, University of Pittsburgh, Pittsburgh, PA 15260\\
	{\bf 2} Fermi National Accelerator Laboratory, Batavia, IL 60510\\
	{\bf 3} Department of Physics, Illinois Institute of Technology, Chicago, IL 60616\\
	{\bf 4} Department of Physics, Southern Methodist University, Dallas, TX 75275\\
	{\bf 5} Jefferson Lab, Newport News, VA 23606\\
	{\bf 6} Department of Physics, College of Sciences, Northeastern University, Shenyang 110819, China\\
	{\bf 7} Department of Physics and Astronomy, Michigan State University, East Lansing, MI 48824\\
	{\bf 8} School of Physics and State Key Laboratory of Nuclear Physics and Technology,\\ 
	Peking University, Beijing 100871, China\\
	$\star$xiekeping@pitt.edu
\end{center}

\begin{center}
\today
\end{center}

\pagestyle{fancy}
\fancyhead[LO]{\colorbox{scipostdeepblue}{\strut \bf \color{white}~Proceedings}}
\fancyhead[RO]{\colorbox{scipostdeepblue}{\strut \bf \color{white}~DIS 2021}}


\definecolor{palegray}{gray}{0.95}
\begin{center}
\colorbox{palegray}{
\begin{tabular}{rr}
\begin{minipage}{0.1\textwidth}
\includegraphics[width=22mm]{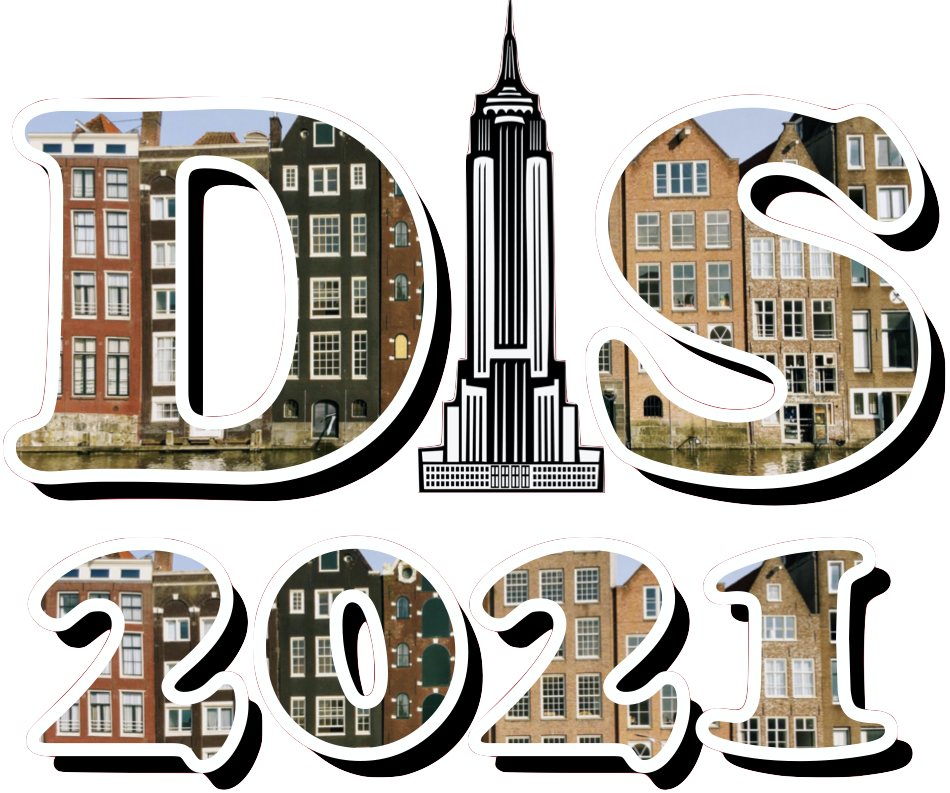}
\end{minipage}
&
\begin{minipage}{0.75\textwidth}
\begin{center}
{\it Proceedings for the XXVIII International Workshop\\ 
   on Deep-Inelastic Scattering and Related Subjects,}\\
{\it Stony Brook University, New York, USA, 12-16 April 2021} \\
\doi{10.21468/SciPostPhysProc.xx.xxxx}
\end{center}
\end{minipage}
\end{tabular}
}
\end{center}

\section*{Abstract}
{
Recently, two photon PDF sets based on implementations of the LUX ansatz into the CT18 global analysis were released.
In CT18lux, the photon PDF is calculated directly using the LUX master formula for all scales, $\mu$. In an alternative realization, CT18qed, the photon PDF is initialized at the starting scale, $\mu_0$, using the LUX formulation and evolved to higher scales $\mu(>\mu_0)$ with a combined QED+QCD kernel at $\mathcal{O}(\alpha),~\mathcal{O}(\alpha\alpha_s)$ and $\mathcal{O}(\alpha^2)$.
In the small-$x$ region, the photon PDF uncertainty is mainly induced by the quark and gluon PDFs, through the perturbative DIS structure functions. In comparison, the large-$x$ photon uncertainty comes from various low-energy, nonperturbative contributions, including variations of the inelastic structure functions in the resonance and continuum regions, higher-twist and target-mass corrections, and elastic electromagnetic form factors of the proton. We take the production of doubly-charged Higgs pairs, $(H^{++}H^{--})$, as an example of scenarios beyond the Standard Model to illustrate the phenomenological implications of these photon PDFs at the LHC.
}


\section{Introduction to the photon PDF}
\label{sec:intro}
With the continued accumulation of experimental data, the Large Hadron Collider (LHC) is increasingly a precision machine. Simultaneously, theoretical calculations have reached next-to-next-to-leading order (NNLO) for many $2\to2$ and some $2\to3$ process, and even next-to-NNLO (N$^{3}$LO) for $2\to1$ processes~\cite{Heinrich:2017una}. At this level of precision, electroweak (EW) corrections begin to have a sizable effect, as $\alpha_e\sim\alpha_S^2$. Today, NLO EW corrections have become standard, and some automatic packages are available~\cite{Biedermann:2017yoi,Frederix:2018nkq}. A consistent NLO EW calculation involves the photon as an active parton of the proton, and photon-initiated (PI) processes can make significant contributions. 

The first photon PDF set incorporating QED corrections to DGLAP evolution was released by the MRST group as MRST2004QED \cite{Martin:2004dh}. In this set, the photon PDF was parameterized as radiation off the ``primordial" up and down quarks, governed by the constituent- and current-quark masses. Alternatively, the NNPDF group included the photon as a new parton, fitting the available high-mass Drell-Yan data in its release of the NNPDF2.3QED~\cite{Ball:2013hta} and NNPDF3.0QED~\cite{Ball:2014uwa} PDFs. A complication with this approach comes from the fact that the photon PDF is not well constrained, especially at high-$x$, both because of the small size of the photon-initiated contribution and the large experimental uncertainty. Contemporarily, the CT14QED PDFs~\cite{Schmidt:2015zda} determined the inelastic photon by invoking isolated photon production in deeply-inelastic scattering, $ep\to e\gamma+X$. The elastic photon was included in the CT14QEDinc PDFs using the Equivalent Photon Approximation (EPA) \cite{Budnev:1974de},

Recently, the LUX group introduced the idea that, by viewing the $ep\to eX$ process as electron scattering from the photon field of the proton~\cite{Manohar:2016nzj,Manohar:2017eqh}, the photon PDF can be determined precisely through the proton structure functions, which are directly measured in experiments or perturbatively calculated in QCD. Since this stride, we have seen a second generation of photon PDFs. The NNPDF group incorporated the LUX formalism to initialize its photon PDF at a high scale, $\mu_0=100~\GeV$, then evolved via QED-corrected DGLAP equations both upwards and downwards in $\mu$ in the NNPDF3.1luxQED PDFs~\cite{Bertone:2017bme}. In comparison, the MMHT group took a low initialization scale $(\mu_0=1~\GeV)$ approach and evolved PDFs upwards to obtain the MMHT2015qed set \cite{Harland-Lang:2019pla}.

Along this line, we apply the LUX formalism in the framework of the CT18 global analysis~\cite{Hou:2019efy}. In the first approach, CT18lux~\cite{Xie:2021equ}, we directly calculate the photon PDF with the LUX formula at any scale. Alternatively, in CT18qed~\cite{Xie:2021equ}, the photon PDF is initialized at a low scale, $\mu_0$, with the LUX method, and evolved to higher scales with mixed QED and QCD kernels, up to $\mathcal{O}(\alpha^2)$ and $\mathcal{O}(\alpha\alpha_s)$. For convenience, we will often refer to the former as the LUX approach, while the later as DGLAP evolution. In general, the photon PDFs with both approaches agree in the intermediate-$x$ region, $10^{-3}\lesssim x\lesssim0.3$, while differing in the low- and large-$x$ regions, which will be discussed in more detail below. 

\section{CT18lux vs CT18qed}

As we mentioned in Sec.~\ref{sec:intro}, the photon PDF in CT18lux is fully determined through the LUX master formula~\cite{Manohar:2016nzj,Manohar:2017eqh},
\begin{equation}
\begin{aligned}\label{eq:LUX}
x \gamma\left(x, \mu^{2}\right) &=\frac{1}{2 \pi \alpha\left(\mu^{2}\right)} \int_{x}^{1} \frac{\mathrm{d} z}{z}\left\{\int _ { \frac { x ^ { 2 } m _ { p } ^ { 2 } } { 1 - z } } ^ { \frac { \mu ^ { 2 } } { 1 - z } } \frac { \mathrm { d } Q ^ { 2 } } { Q ^ { 2 } } \alpha _ { \mathrm { ph } } ^ { 2 } ( - Q ^ { 2 } ) \Bigg[\left(z p_{\gamma q}(z)+\frac{2 x^{2} m_{p}^{2}}{Q^{2}}\right) F_{2}\left(x / z, Q^{2}\right)\right.\\
&\left.-z^{2} F_{L}\left(x / z, Q^{2}\right)\Bigg]-\alpha^{2}\left(\mu^{2}\right) z^{2} F_{2}\left(x / z, \mu^{2}\right)\right\}+\mathcal{O}\left(\alpha^{2}, \alpha \alpha_{s}\right).
\end{aligned}
\end{equation}
The integrated, $\int\dd Q^2/Q^2$, square-bracket term above is designated the physical factorization term, while the rest involving the negative of $F_2$ is the $\overline{\rm MS}$ conversion term.
In comparison, the CT18qed photon is evolved according to the DGLAP equations,
\begin{equation}\label{eq:DGLAP}
\frac{\mathrm{d} \gamma}{\mathrm{d} \log \mu^{2}}=\frac{\alpha}{2 \pi}\left[p_{\gamma \gamma} \otimes \gamma+\sum_{i} e_{i}^{2} p_{\gamma q} \otimes\left(q_{i}+\bar{q}_{i}\right)\right].
\end{equation}

\begin{figure}\centering
\includegraphics[width=0.48\textwidth]{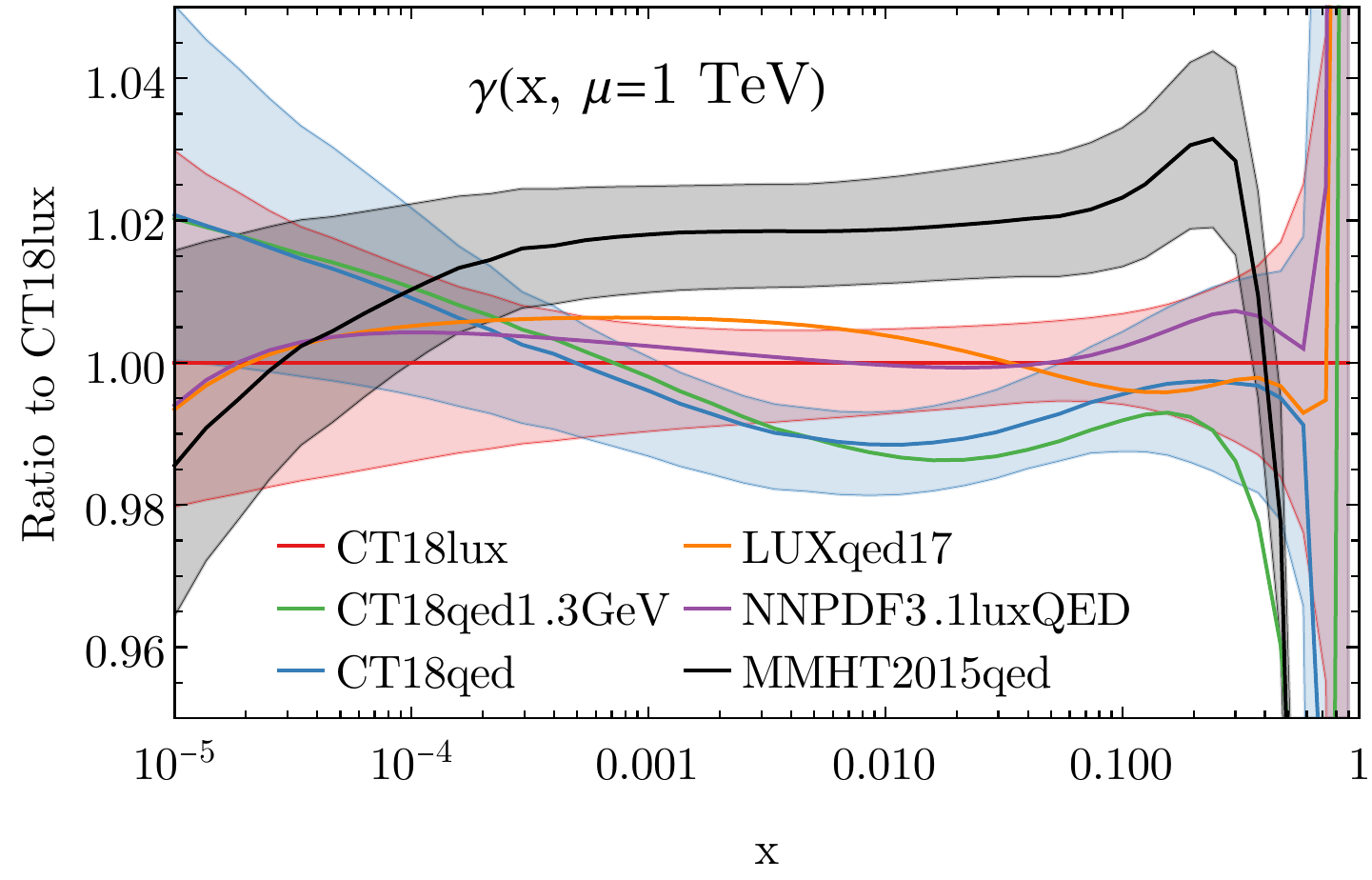}
\includegraphics[width=0.48\textwidth]{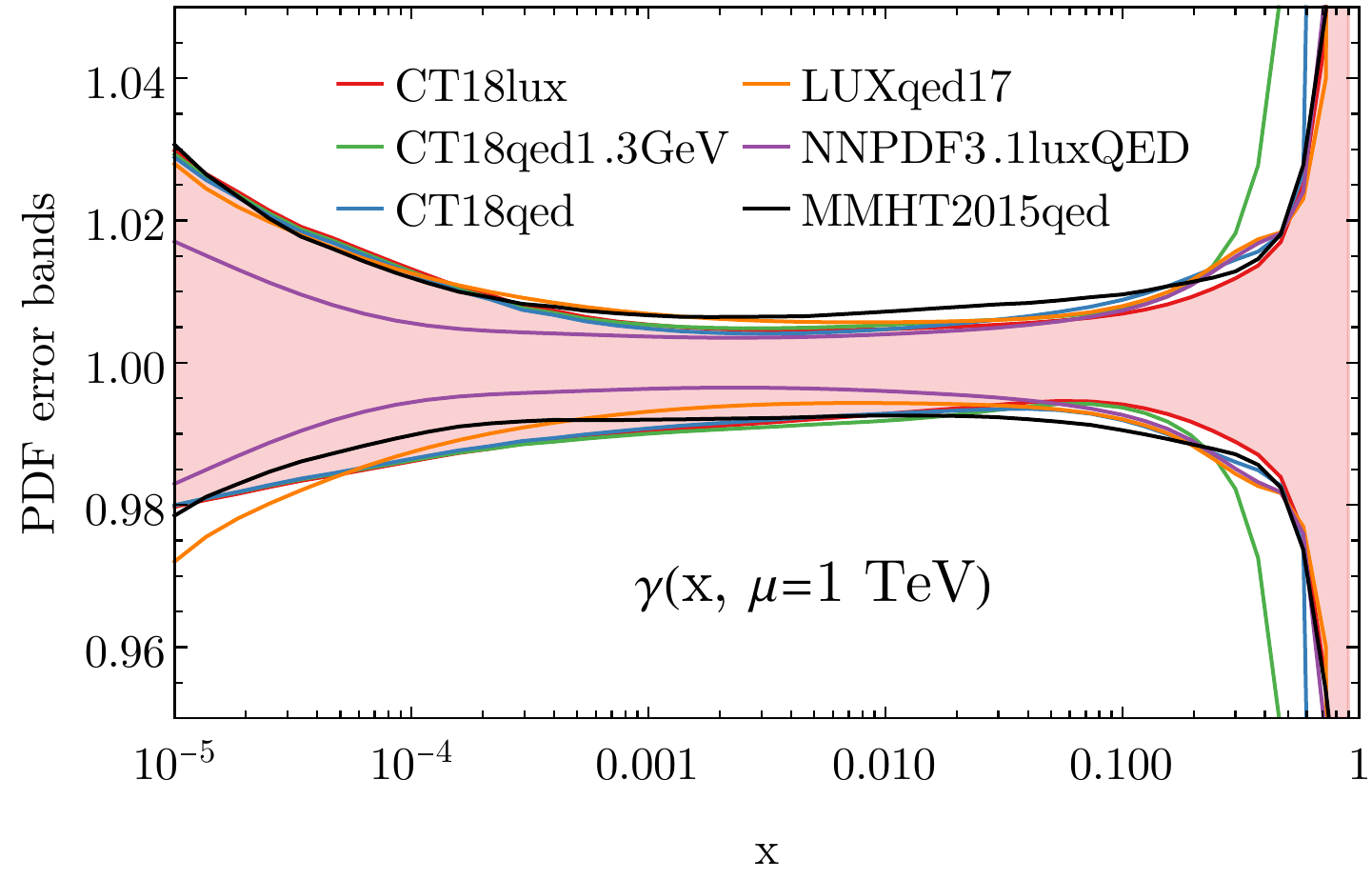}
\caption{Comparisons of the second-generation photon PDFs and their corresponding error bands at $\mu=1$ TeV.}
\label{fig:photon1TeV}
\end{figure}
A comparison of the photon PDFs in the EW precision region at $\mu=100~\GeV$ is shown in Ref.~\cite{Xie:2021equ}. In these proceedings, we extend this comparison to $\mu=1~\TeV$ in Fig.~\ref{fig:photon1TeV}, which probes physics beyond the Standard Model (BSM). Similarly as before, we see that CT18qed gives an enhancement to the photon PDF in the low-$x$ region as the DGLAP evolution in Eq.~(\ref{eq:DGLAP}) gives an equivalent leading-order structure function, $F_{2}^{\rm LO}=x\sum_{i}e_i^2(q_i+\bar{q}_{i})$\footnote{At LO, \emph{i.e.}, $\mathcal{O}(\alpha_s^0)$, $F_L=0$. }, to the LUX formulation, Eq.~(\ref{eq:LUX}). In comparison, the LUX approach incorporates the full proton structure functions, calculated at NNLO in the pQCD region and smaller than $F_{2}^{\rm LO}$ \cite{Xie:2021equ}. 
At large $x$, the DGLAP approach gives a significantly smaller photon than the LUX approach, due to the large nonperturbative structure functions in the negative $\overline{\rm MS}$ conversion term.  

In CT18qed, two initialization scales are explored: $\mu_0=1.3~\GeV$, which is within the low-$Q^2$ nonperturbative region, and $\mu_0=3~\GeV$, in the perturbative QCD region. We recommend $\mu_0=3~\GeV$ as the nominal set for CT18qed due to its smaller uncertainty in the large-$x$ region, shown in the right plot of Fig.~\ref{fig:photon1TeV}. The lower initialization scale, $\mu_0=1.3~\GeV$, same as the default choice for CT18~\cite{Hou:2019efy}, is more appropriate for describing the photon PDF in the low-energy region. 

\begin{figure}\centering
	\includegraphics[width=0.48\textwidth]{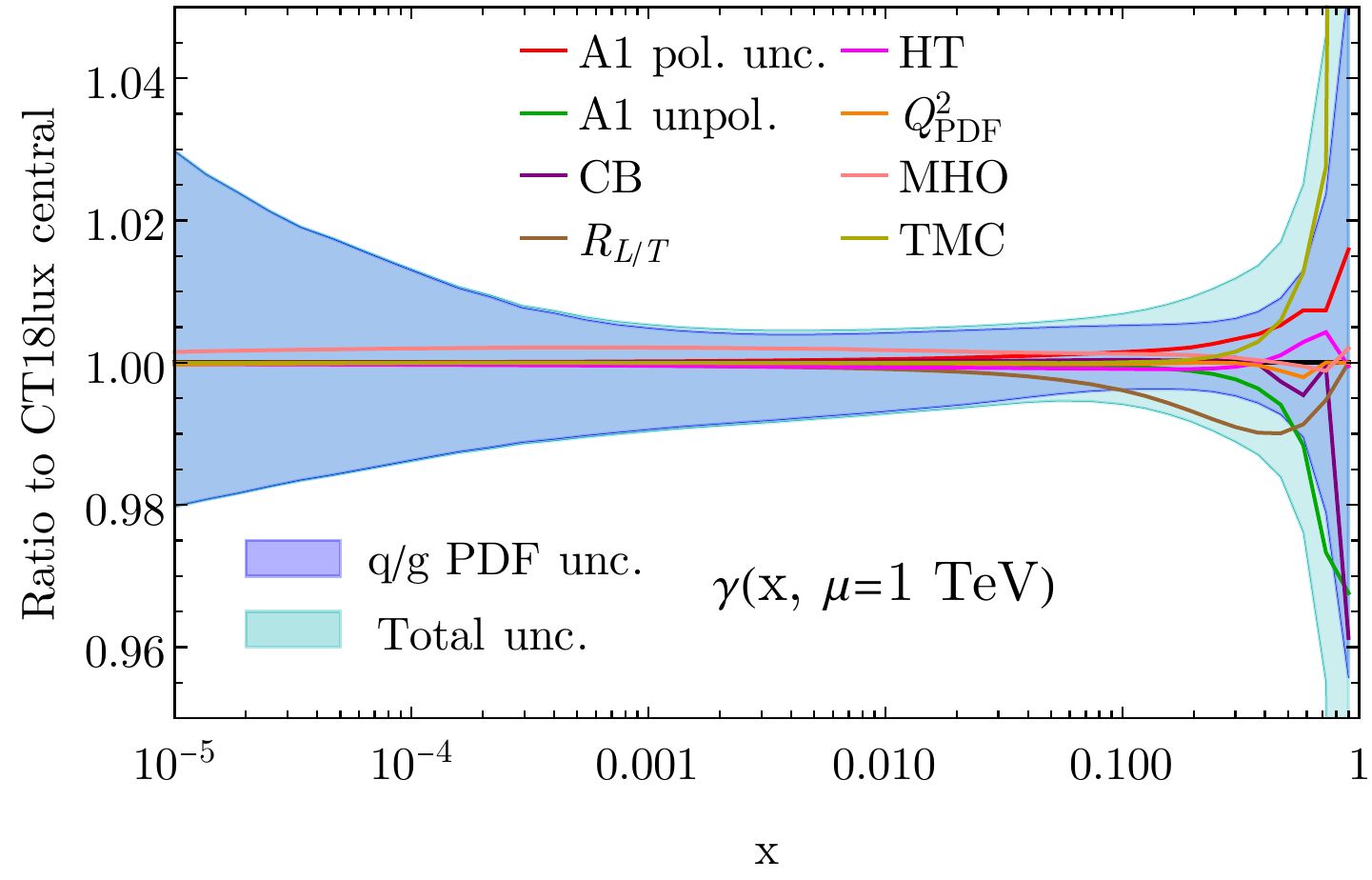}
	\includegraphics[width=0.48\textwidth]{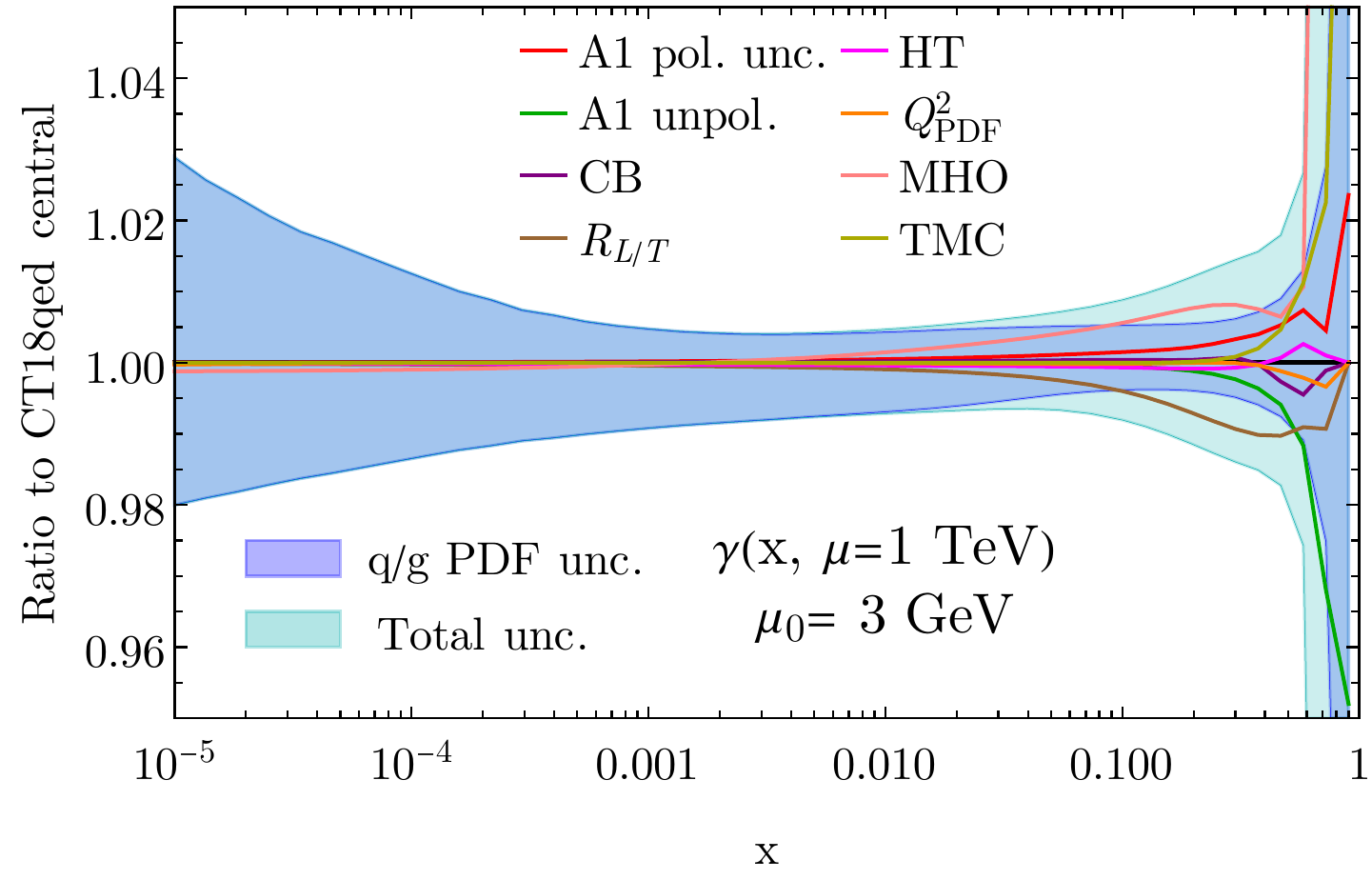}
	\caption{The individual contributions to the full $\gamma$-PDF uncertainty in the CT18lux and CT18qed (with $\mu_0=3$ GeV) calculations. The various sources of uncertainty are added on top
	of the uncertainties associated with variations of the quark- and gluon-PDF parameters in the LUX master formula.}
	\label{fig:unc}
\end{figure}
Similarly to Ref.~\cite{Xie:2021equ}, we have explored various sources contributing to the photon PDF uncertainty in Fig.~\ref{fig:unc}.
We see that in the low-$x$ region the photon uncertainty is mainly induced by the quark-gluon PDFs, through the perturbative structure functions in the high-$Q^2$ pQCD region. In comparison, the large-$x$ uncertainty is dominated by various low-$Q^2$ contributions: the statistical and model errors of the A1 polarized fit of the elastic form factors (A1 pol.~unc.)~\cite{Bernauer:2013tpr}; the variation of the A1 fit of the unpolarized data (A1 unpol.)~\cite{Bernauer:2013tpr}; the effect of changing the resonance structure functions from the CLAS fit~\cite{Osipenko:2003bu} to the Christy-Bosted fit (CB)~\cite{Christy:2007ve,Christy:2021abc}; conservatively assigning a $\pm50\%$ uncertainty to the HERMES measurement of $R_{L/T}=\sigma_L/\sigma_T$~\cite{Airapetian:2011nu}; adding a higher-twist (HT) correction~\cite{Accardi:2016qay,Abt:2016vjh}; changing the matching scale, $Q^2_{\rm PDF}$, between the low-$Q^2$ nonperturbative and high-$Q^2$ pQCD regions; probing the missing higher-order (MHO) effect by varying the separation scale, $M^2[z]=\mu^2/(1-z)$, in Eq.~(\ref{eq:LUX}) to $\mu^2$; and target-mass corrections (TMC)~\cite{Schienbein:2007gr,Brady:2011uy}. 

\section{A phenomenological application: $H^{\pm\pm}$ pair production}
\begin{figure}
	\centering
	\includegraphics[width=0.9\textwidth]{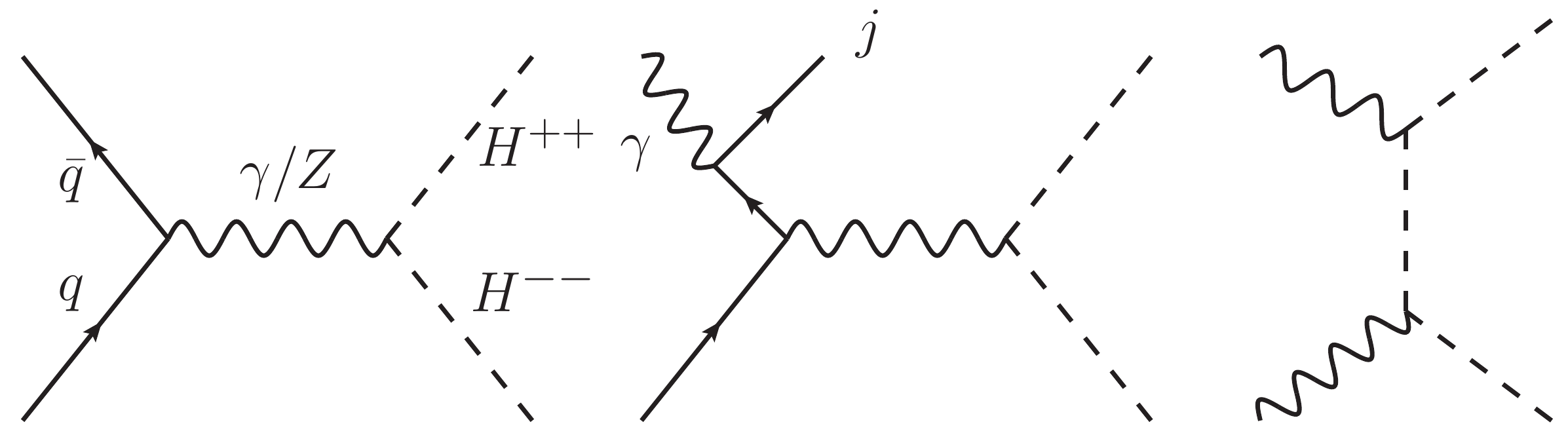}
	\caption{Representative Feynman diagrams for $H^{\pm\pm}$ pair production through Drell-Yan-like, single-, and double-photon initiated processes.}
	\label{fig:feynman}
\end{figure}

Applications of photon PDFs to several SM processes have already been explored in Ref.~\cite{Xie:2021equ}. In these proceedings, we take doubly-charged Higgs $H^{\pm\pm}$ pair production as an important example to extend in this direction further to a specific BSM scenario. Representative Feynman diagrams are illustrated in Fig.~\ref{fig:feynman}. The well-known mechanism for the $H^{\pm\pm}$ pair production is QCD Drell-Yan-like quark-antiquark annihilation. At EW NLO, single-photon initiated (SPI) processes emerge, shown in Fig.~\ref{fig:feynman}(b). In addition, due to the large electric charge, we would also expect the production rate of the double-photon initiated (DPI) processes to get an enhancement of a factor of $2^4=16$, compared to singly-charged particle pair production. 

\begin{figure}\centering
	\includegraphics[width=0.52\textwidth]{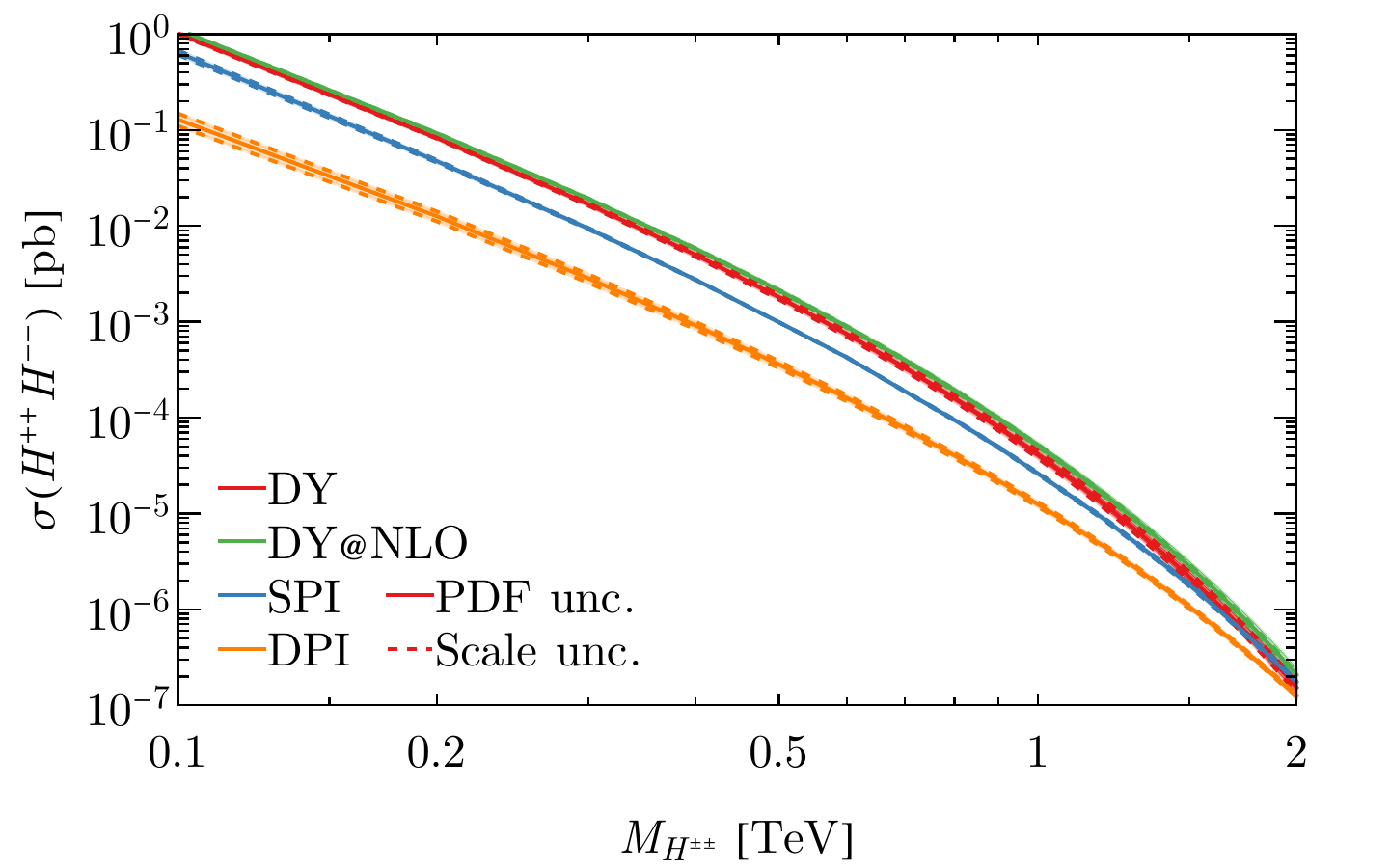}
	\includegraphics[width=0.47\textwidth]{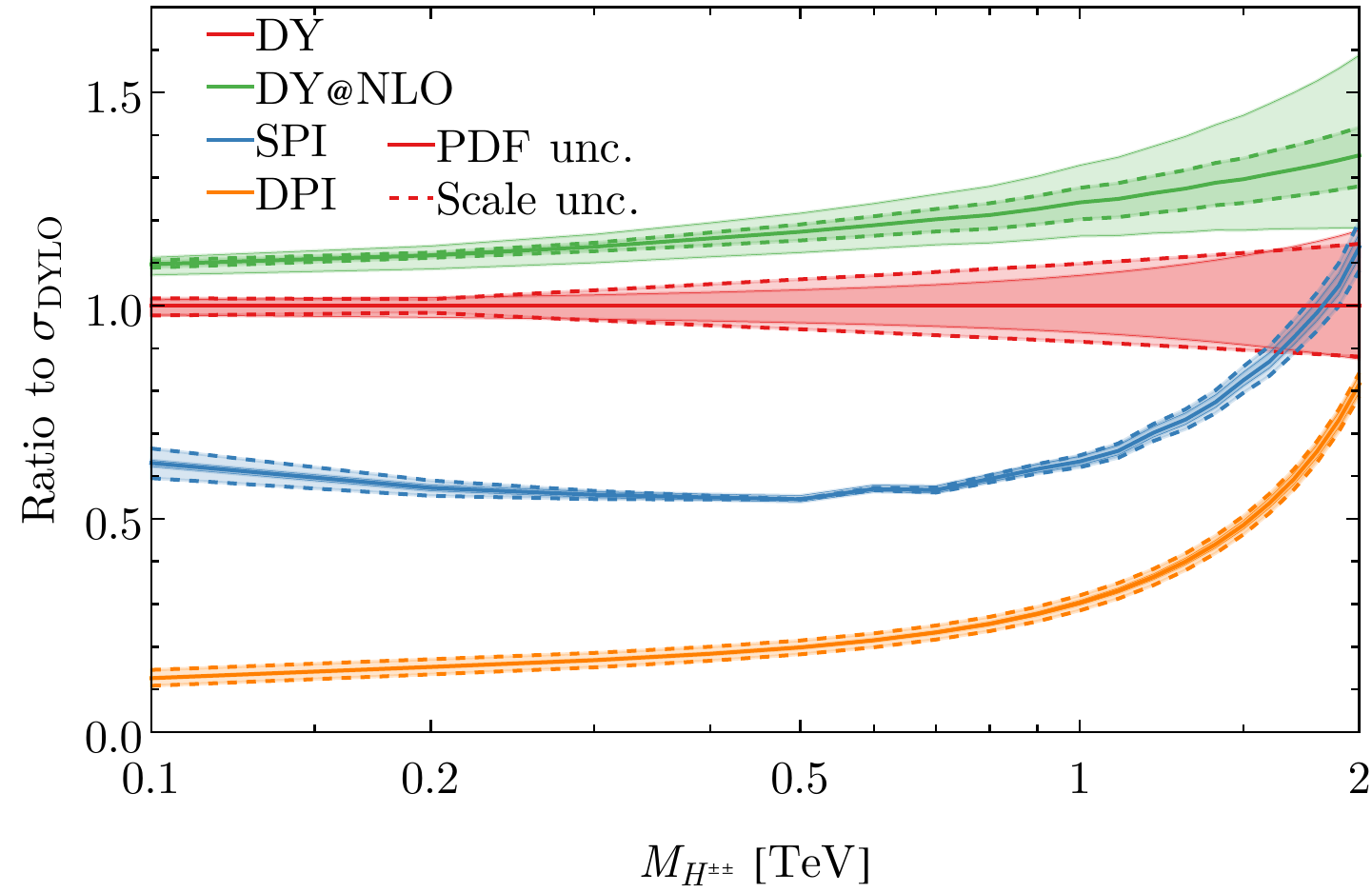}
	\includegraphics[width=0.485\textwidth]{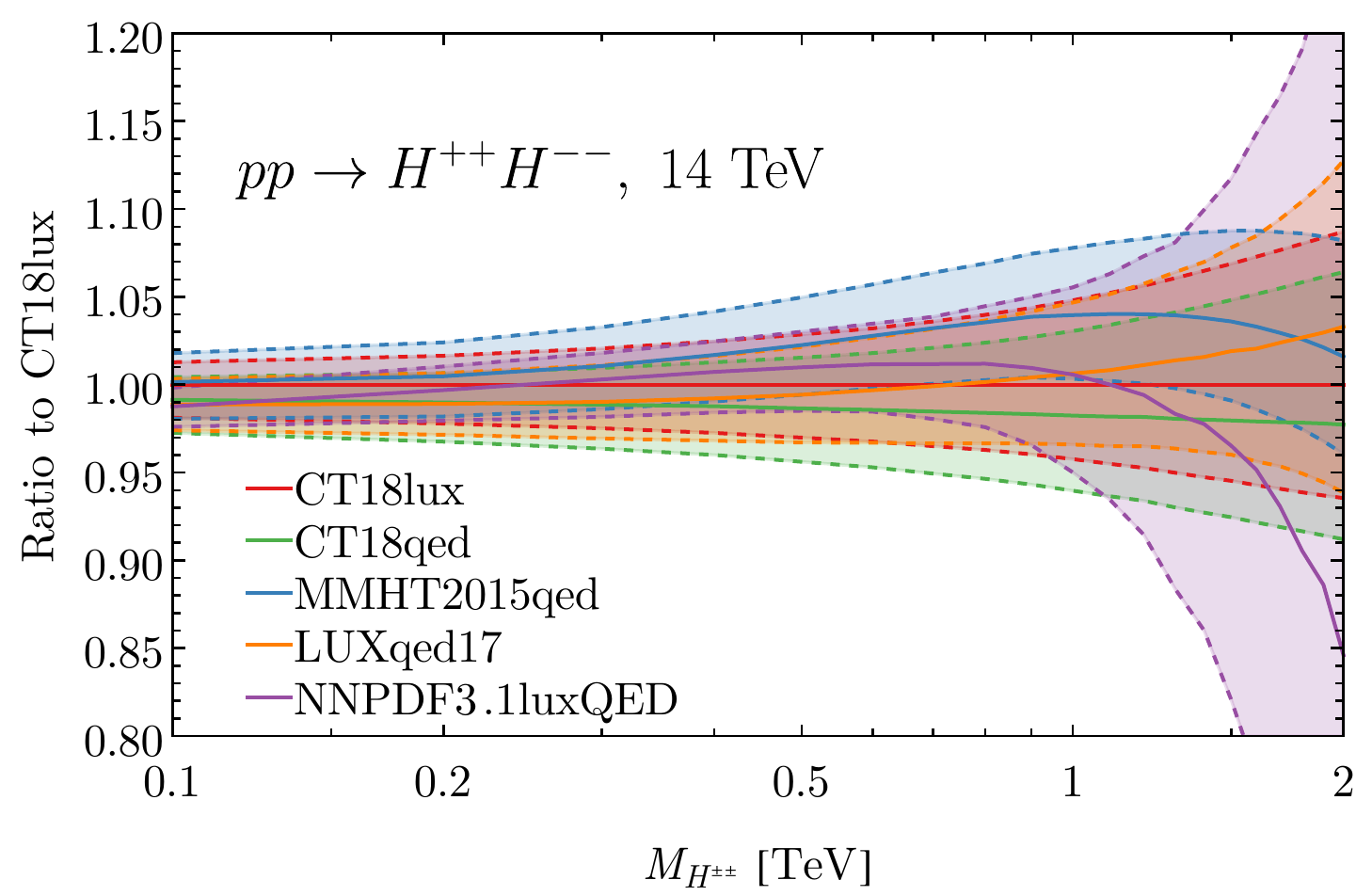}
	\includegraphics[width=0.505\textwidth]{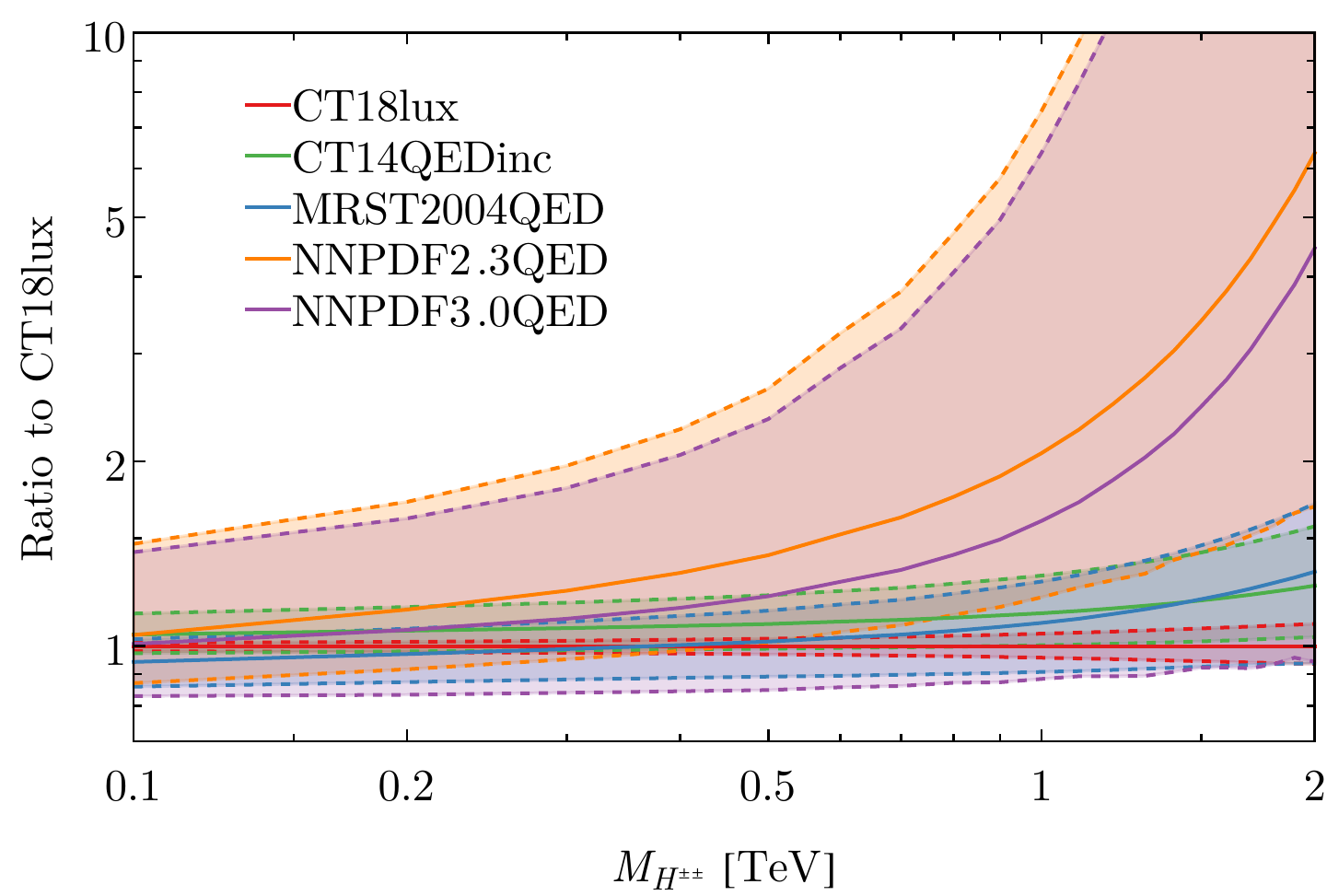}	
	\caption{The cross sections for $H^{\pm\pm}$ pair production at a 14 TeV proton-proton collider.}
	\label{fig:xsec}
\end{figure}
The total inclusive\footnote{Here the ``total inclusive" means the full phase space for the production of on-shell $H^{\pm\pm}$ pairs without decay.} cross sections for $H^{\pm\pm}$ pair production at a 14 TeV proton-proton collider through different mechanisms versus the $H^{\pm\pm}$ mass are shown in Fig.~\ref{fig:xsec}. The corresponding ratios normalized to the LO DY cross sections are shown in the upper-right panel of Fig.~\ref{fig:xsec}. We see the NLO QCD corrections give a $K$-factor of about $1.1\sim1.35$ when the Higgs mass increases from 100 GeV to 2 TeV. The PDF uncertainty increases from 2\% to 15\%.
We have also estimated the scale uncertainty with the 7-point approach by varying the renormalization and factorization scales as
\begin{equation}
(\mu_R,\mu_F)=\{(1/2,1/2),(1/2,1),(1,1/2),(1,1),(1,2),(2,1),(2,2)\}\sqrt{\hat{s}},
\end{equation}
where $\sqrt{\hat{s}}$ is partonic energy. As shown in Fig.~\ref{fig:xsec}(b), the scale uncertainty for the LO DY cross sections is roughly the same size as the PDF uncertainty, but is reduced significantly by a factor of $2\sim3$ at NLO. 

Compared with the Drell-Yan mechanism, the single-photon initiated processes make about a $60\%$ contribution at low $H^{\pm\pm}$ mass, while increasing up to the same size at $M_{H^{\pm\pm}}=2~\TeV$. The contribution of double-photon initiated processes increases from 10\% to 80\% when compared with LO DY, highlighting the importance of the photon contribution. 

The PDF uncertainty of the total cross section from summing the NLO DY, SPI and DPI contributions is shown in the lower two plots of Fig.~\ref{fig:xsec}. In general, the PDF uncertainty of the total cross section increases from 2\% to 10\% for CT18lux and CT18qed. In comparison, LUXqed17 gives a slightly larger error band, as the quark and gluon PDFs were taken from the PDF4LHC15 set~\cite{Butterworth:2015oua}, which was based on a previous round of global fits. The CT18lux and CT18qed uncertainties are slightly larger than MMHT2015qed's, as a new low-$Q^2$ uncertainty source, the target-mass correction, was included as an individual error set.
Intriguingly, NNPDF3.1luxQED gives a significantly larger PDF uncertainty in the large $M_{H^{\pm\pm}}$ tail than the other sets, due to the larger extrapolation error bands for the quark partons in the large-$x$ region when $x\to1$; this induces a larger error band for the Drell-Yan cross section in this region.

In the lower-right plot of Fig.~\ref{fig:xsec}, we compare the PDF uncertainty from the first generation of photon PDFs with CT18lux. 
As expected, the NNPDF2.3QED and NNPDF3.0QED PDFs are significantly larger than the others, because of the diminished constraints from high-mass Drell Yan data. 
CT14QEDinc gives overall agreement with CT18lux, although with a larger error band. 
MRST2004QED gives smaller central predictions at low $M_{H^{\pm\pm}}$ when compared with CT18lux, but gradually exceeds the latter as the $H^{\pm\pm}$ mass increases.

\section{Summary}
In these proceedings, we summarized the main development of the recent two photon PDF sets, CT18lux and CT18qed, released by the CTEQ-TEA group~\cite{Xie:2021equ}. In CT18lux, we calculated the photon PDF in terms of the LUX formalism at all scales, while the CT18qed evolves the photon together with other partons in terms of QED-corrected DGLAP equations, up to $\mathcal{O}(\alpha\alpha_s)$ and $\mathcal{O}(\alpha^2)$. The two different approaches are in overall agreement for the photon PDF in the intermediate region. In contrast, the CT18qed gives a larger photon at low $x$, and significantly smaller at large $x$. We take the $H^{\pm\pm}$ pair production as a demonstration of the application of photon PDFs to BSM physics. We find the single- and double-photon initiated processes make significant contributions to the total cross section for doubly-charged Higgs pair production. Compared with the first generation, all second-generation photon PDFs based on the LUX approach improve precision, generally up to the percent-level. 

\section*{Acknowledgment}
We thank Eric Christy for providing updated code to compute the low-$W^2$ resonance-region structure functions, as well as for valuable conversations. We also thank our the CTEQ-TEA colleagues for helpful discussions and Sergei Kulagin for useful exchanges. 
The work at MSU is partially supported by the U.S.~National Science Foundation
under Grant No.~PHY-2013791.
The work of T.~J.~Hobbs was supported by the U.S.~Department of
Energy under Grant No.~DE-SC0010129 as well as by a JLab EIC Center Fellowship.
The work of K.~Xie was supported in part 
by the U.S.~Department of Energy under
grant No.~DE-FG02-95ER40896, U.S.~National Science Foundation under Grant No.~PHY-1820760, and in part by the PITT PACC.
The work of M.~Yan is supported by the National Science Foundation of China under Grant Nos. 11725520, 11675002, and 11635001.
C.-P.~Yuan is also grateful for the support from
the Wu-Ki Tung endowed chair in particle physics.
Work supported by the Fermi National Accelerator Laboratory, managed and operated by Fermi Research Alliance, LLC under Contract No.~DE-AC02-07CH11359 with the U.S.~Department of Energy.

\bibliography{ref_Xie.bib}

\begin{thebibliography}{10}
\providecommand{\url}[1]{\texttt{#1}}
\providecommand{\urlprefix}{URL }
\expandafter\ifx\csname urlstyle\endcsname\relax
  \providecommand{\doi}[1]{doi:\discretionary{}{}{}#1}\else
  \providecommand{\doi}{doi:\discretionary{}{}{}\begingroup
  \urlstyle{rm}\Url}\fi
\providecommand{\eprint}[2][]{\url{#2}}

\bibitem{Heinrich:2017una}
G.~Heinrich,
\newblock \emph{{QCD calculations for the LHC: status and prospects}},
\newblock In \emph{{5th Large Hadron Collider Physics Conference}} (2017),
  \eprint{1710.04998}.

\bibitem{Biedermann:2017yoi}
B.~Biedermann, S.~Br\"auer, A.~Denner, M.~Pellen, S.~Schumann and J.~M.
  Thompson,
\newblock \emph{{Automation of NLO QCD and EW corrections with Sherpa and
  Recola}},
\newblock Eur. Phys. J. C \textbf{77}, 492 (2017),
\newblock \doi{10.1140/epjc/s10052-017-5054-8},
\newblock \eprint{1704.05783}.

\bibitem{Frederix:2018nkq}
R.~Frederix, S.~Frixione, V.~Hirschi, D.~Pagani, H.~S. Shao and M.~Zaro,
\newblock \emph{{The automation of next-to-leading order electroweak
  calculations}},
\newblock JHEP \textbf{07}, 185 (2018),
\newblock \doi{10.1007/JHEP07(2018)185},
\newblock \eprint{1804.10017}.

\bibitem{Martin:2004dh}
A.~Martin, R.~Roberts, W.~Stirling and R.~Thorne,
\newblock \emph{{Parton distributions incorporating QED contributions}},
\newblock Eur. Phys. J. C \textbf{39}, 155 (2005),
\newblock \doi{10.1140/epjc/s2004-02088-7},
\newblock \eprint{hep-ph/0411040}.

\bibitem{Ball:2013hta}
R.~D. Ball, V.~Bertone, S.~Carrazza, L.~Del~Debbio, S.~Forte, A.~Guffanti,
  N.~P. Hartland and J.~Rojo,
\newblock \emph{{Parton distributions with QED corrections}},
\newblock Nucl. Phys. B \textbf{877}, 290 (2013),
\newblock \doi{10.1016/j.nuclphysb.2013.10.010},
\newblock \eprint{1308.0598}.

\bibitem{Ball:2014uwa}
R.~D. Ball \emph{et~al.},
\newblock \emph{{Parton distributions for the LHC Run II}},
\newblock JHEP \textbf{04}, 040 (2015),
\newblock \doi{10.1007/JHEP04(2015)040},
\newblock \eprint{1410.8849}.

\bibitem{Schmidt:2015zda}
C.~Schmidt, J.~Pumplin, D.~Stump and C.~Yuan,
\newblock \emph{{CT14QED parton distribution functions from isolated photon
  production in deep inelastic scattering}},
\newblock Phys. Rev. D \textbf{93}(11), 114015 (2016),
\newblock \doi{10.1103/PhysRevD.93.114015},
\newblock \eprint{1509.02905}.

\bibitem{Budnev:1974de}
V.~M. Budnev, I.~F. Ginzburg, G.~V. Meledin and V.~G. Serbo,
\newblock \emph{{The Two photon particle production mechanism. Physical
  problems. Applications. Equivalent photon approximation}},
\newblock Phys. Rept. \textbf{15}, 181 (1975),
\newblock \doi{10.1016/0370-1573(75)90009-5}.

\bibitem{Manohar:2016nzj}
A.~Manohar, P.~Nason, G.~P. Salam and G.~Zanderighi,
\newblock \emph{{How bright is the proton? A precise determination of the
  photon parton distribution function}},
\newblock Phys. Rev. Lett. \textbf{117}(24), 242002 (2016),
\newblock \doi{10.1103/PhysRevLett.117.242002},
\newblock \eprint{1607.04266}.

\bibitem{Manohar:2017eqh}
A.~V. Manohar, P.~Nason, G.~P. Salam and G.~Zanderighi,
\newblock \emph{{The Photon Content of the Proton}},
\newblock JHEP \textbf{12}, 046 (2017),
\newblock \doi{10.1007/JHEP12(2017)046},
\newblock \eprint{1708.01256}.

\bibitem{Bertone:2017bme}
V.~Bertone, S.~Carrazza, N.~P. Hartland and J.~Rojo,
\newblock \emph{{Illuminating the photon content of the proton within a global
  PDF analysis}},
\newblock SciPost Phys. \textbf{5}(1), 008 (2018),
\newblock \doi{10.21468/SciPostPhys.5.1.008},
\newblock \eprint{1712.07053}.

\bibitem{Harland-Lang:2019pla}
L.~Harland-Lang, A.~Martin, R.~Nathvani and R.~Thorne,
\newblock \emph{{Ad Lucem: QED Parton Distribution Functions in the MMHT
  Framework}},
\newblock Eur. Phys. J. C \textbf{79}(10), 811 (2019),
\newblock \doi{10.1140/epjc/s10052-019-7296-0},
\newblock \eprint{1907.02750}.

\bibitem{Hou:2019efy}
T.-J. Hou \emph{et~al.},
\newblock \emph{{New CTEQ global analysis of quantum chromodynamics with
  high-precision data from the LHC}},
\newblock Phys. Rev. D \textbf{103}(1), 014013 (2021),
\newblock \doi{10.1103/PhysRevD.103.014013},
\newblock \eprint{1912.10053}.

\bibitem{Xie:2021equ}
K.~Xie, T.~J. Hobbs, T.-J. Hou, C.~Schmidt, M.~Yan and C.~P. Yuan,
\newblock \emph{{The photon PDF within the CT18 global analysis}}  (2021),
\newblock \eprint{2106.10299}.

\bibitem{Bernauer:2013tpr}
J.~C. Bernauer \emph{et~al.},
\newblock \emph{{Electric and magnetic form factors of the proton}},
\newblock Phys. Rev. C \textbf{90}(1), 015206 (2014),
\newblock \doi{10.1103/PhysRevC.90.015206},
\newblock \eprint{1307.6227}.

\bibitem{Osipenko:2003bu}
M.~Osipenko \emph{et~al.},
\newblock \emph{{A Kinematically complete measurement of the proton structure
  function F(2) in the resonance region and evaluation of its moments}},
\newblock Phys. Rev. D \textbf{67}, 092001 (2003),
\newblock \doi{10.1103/PhysRevD.67.092001},
\newblock \eprint{hep-ph/0301204}.

\bibitem{Christy:2007ve}
M.~Christy and P.~E. Bosted,
\newblock \emph{{Empirical fit to precision inclusive electron-proton cross-
  sections in the resonance region}},
\newblock Phys. Rev. C \textbf{81}, 055213 (2010),
\newblock \doi{10.1103/PhysRevC.81.055213},
\newblock \eprint{0712.3731}.

\bibitem{Christy:2021abc}
M.~E. Christy, N.~Kalantarians, J.~Either and W.~Melnitchouk,
\newblock \emph{{To be published}}  (2021).

\bibitem{Airapetian:2011nu}
A.~Airapetian \emph{et~al.},
\newblock \emph{{Inclusive Measurements of Inelastic Electron and Positron
  Scattering from Unpolarized Hydrogen and Deuterium Targets}},
\newblock JHEP \textbf{05}, 126 (2011),
\newblock \doi{10.1007/JHEP05(2011)126},
\newblock \eprint{1103.5704}.

\bibitem{Accardi:2016qay}
A.~Accardi, L.~Brady, W.~Melnitchouk, J.~Owens and N.~Sato,
\newblock \emph{{Constraints on large-$x$ parton distributions from new weak
  boson production and deep-inelastic scattering data}},
\newblock Phys. Rev. D \textbf{93}(11), 114017 (2016),
\newblock \doi{10.1103/PhysRevD.93.114017},
\newblock \eprint{1602.03154}.

\bibitem{Abt:2016vjh}
I.~Abt, A.~M. Cooper-Sarkar, B.~Foster, V.~Myronenko, K.~Wichmann and M.~Wing,
\newblock \emph{{Study of HERA ep data at low Q$^2$ and low $x_{Bj}$ and the
  need for higher-twist corrections to standard perturbative QCD fits}},
\newblock Phys. Rev. D \textbf{94}(3), 034032 (2016),
\newblock \doi{10.1103/PhysRevD.94.034032},
\newblock \eprint{1604.02299}.

\bibitem{Schienbein:2007gr}
I.~Schienbein \emph{et~al.},
\newblock \emph{{A Review of Target Mass Corrections}},
\newblock J. Phys. G \textbf{35}, 053101 (2008),
\newblock \doi{10.1088/0954-3899/35/5/053101},
\newblock \eprint{0709.1775}.

\bibitem{Brady:2011uy}
L.~Brady, A.~Accardi, T.~Hobbs and W.~Melnitchouk,
\newblock \emph{{Next-to leading order analysis of target mass corrections to
  structure functions and asymmetries}},
\newblock Phys. Rev. D \textbf{84}, 074008 (2011),
\newblock \doi{10.1103/PhysRevD.84.074008},
\newblock [Erratum: Phys.Rev.D 85, 039902 (2012)],
\newblock \eprint{1108.4734}.

\bibitem{Butterworth:2015oua}
J.~Butterworth \emph{et~al.},
\newblock \emph{{PDF4LHC recommendations for LHC Run II}},
\newblock J. Phys. G \textbf{43}, 023001 (2016),
\newblock \doi{10.1088/0954-3899/43/2/023001},
\newblock \eprint{1510.03865}.

\end{thebibliography}

\end{document}